# Manipulating the metal-insulator transitions in correlated vanadium dioxide through bandwidth and band-filling control

*Xiaohui Yao [1], Jiahui Ji [1], Xuanchi Zhou [1, 2] \**

[1] *Key Laboratory of Magnetic Molecules and Magnetic Information Materials of Ministry of Education & School of Materials Science and Engineering, Shanxi Normal University, Taiyuan, 030031, China*
[2] *Research Institute of Materials Science, Shanxi Key Laboratory of Advanced Magnetic Materials and Devices, Shanxi Normal University, Taiyuan 030031, China*

*Authors to whom correspondence should be addressed: xuanchizhou@sxnu.edu.cn (X. Zhou).*

**Abstract**

The metal-insulator transition (MIT) in correlated oxide systems opens up a new paradigm to trigger the abruption in multiple physical functionalities, enabling the possibility in unlocking exotic quantum states beyond conventional phase diagram. Nevertheless, the critical challenge for practical device implementation lies in achieving the precise control over the MIT behavior of correlated system across a broad temperature range, ensuring the operational adaptability in diverse environments. Herein, correlated vanadium dioxide ($VO_2$) serves as a model system to demonstrate effective modulations on the MIT functionality through bandwidth and band-filling control. Leveraging the lattice mismatching between $RuO_2$ buffer layer and $TiO_2$ substrate, the *in-plane* tensile strain states in $VO_2$ films can be continuously adjusted by simply altering the thickness of buffer layer, leading to a tunable MIT property over a wide range exceeding 20 K. Beyond that, proton evolution is unveiled to drive the structural transformation of $VO_2$, with a pronounced strain dependence, which is accompanied by hydrogenation-triggered collective carrier delocalization through hydrogen-related band filling in $t_{2g}$ band. The present work establishes an enticing platform for tailoring the MIT properties in correlated electron systems, paving the way for the rational design in exotic electronic phases and physical phenomena.

## 1. Introduction

The complex interactions between the charge, lattice, orbital, and spin degrees of freedom in correlated oxide systems enable the possibility in exploring exotic physical functionality and phenomena, for example, superconductivity,[1, 2] metal-insulator transition (MIT) [3-5] and ferroics.[6-8] In particular, the MIT functionality, featured by the transition from itinerant to localized electronic behaviour, enables synergistic modulation on multiple physical properties, which is harnessed in correlated electronics,[9-11] neuromorphic learning [12, 13] and thermochromic.[14, 15] Among the existing material systems, vanadium dioxide ($VO_2$) undergoes the most abrupt MIT behavior near room temperature that is accompanied by a monoclinic-to-rutile structural transformation, attractive for device applications.[16] The underlying mechanism of the MIT in $VO_2$ remains debated, centering on whether the phase transition is driven primarily by V-V dimerization or by the splitting of the $d_{//}$ orbital.[17, 18] However, the critical factor facilitating practical device applications lies in the realization of a widely tunable MIT functionality across a broad temperature range, catering for operational adaptability in diverse environments. Beyond traditional semiconductors, correlated oxides exhibit exceptionally sensitive electronic orbital configurations that can be readily adjusted using chemical doping,[19, 20] strain engineering,[21, 22] and oxygen defects,[23] particularly for those near a competing metal-insulator phase boundary.

From the perspective of Mott physics, the MIT behaviors in correlated oxides can be effectively regulated through two primary perspectives: 1) bandwidth control via modifying the electron correlation strength relative to the kinetic energy; 2) band-filling control through directly tailoring the carrier density and orbital filling. Specially, imparting the high pressure to $VO_2$ material results in novel crystalline structures (e.g., M1', R, O and X metastable phases) with collective metallization driven by a wider bandwidth.[16, 24] Beyond that, hydrogenation offers an alternative pathway for manipulating the MIT property in correlated $VO_2$ through band-filling control, as each incorporated hydrogen can donate one electron into the conduction band.[25-27] As a result, hydrogen-associated electron doping process directly adjusts the $d$-orbital occupancy and configuration of $VO_2$, triggering Mott phase transitions away from correlated electron ground state based on $t_{2g}^1 e_g^0$ configuration.[28] Similarly, substituting tetravalent vanadium of $VO_2$ with high-valence $W^{6+}$ dopants reduces the transition temperature ($T_{MIT}$) towards room temperature through electron doping, satisfying the room-temperature operation for smart windows.[14] Therefore, realizing the precise control over the MIT property of $VO_2$ is rather critical for both the Mott physics and correlated electronic device applications.

In this work, we showcase a synergistic strategy for the precise control over the MIT functionality of $VO_2$ through bandwidth and band-filling control. As a representative case, the *in-plane* interfacial strain states in $VO_2$ films deposited on *c*-plane $TiO_2$ templates can be facilely adjusted by engineering the thickness of $RuO_2$ buffer layer, thus rendering a continuous modulation on the $T_{MIT}$ over a wide temperature range. Benefiting from the proton evolution, the MIT functionality of $VO_2$ can be further depressed via hydrogen-associated electron doping, with a pronounced lattice expansion correlated with the strain states in $VO_2$. The present work offers a fertile ground for manipulating the MIT properties in electron-correlated system, enabling the rational design in novel electronic phases unattainable in conventional phase diagram.

## 2. Experimental Section

The $VO_2$ films were grown on the single crystalline *c*-plane $TiO_2$ substrates through the laser molecular beam epitaxy (LMBE) technique, with the $RuO_2$ delicately selected as the buffer layer. Before the film deposition, the primary chamber was evacuated to a base pressure below $10^{-5}$ Pa. The film deposition was carried out at 400 °C under an oxygen partial pressure of 1.5 Pa, with a target-substrate distance of 45 mm and a laser fluence of 1.0 J·cm$^{-2}$. Afterwards, the as-deposited $VO_2/RuO_2/TiO_2$ (001) heterostructures were naturally cooled down to the room temperature under an identical oxygen partial pressure to the film deposition. Prior to the hydrogenation, the 20 nm-thick platinum dots were sputtered into the surface of as-grown $VO_2/RuO_2/TiO_2$ (001) heterostructures using magnetron sputtering. Finally, hydrogenation was achieved through a hydrogen spillover approach by annealing the $Pt/VO_2/RuO_2/TiO_2$ (001) heterostructures in a 5 % $H_2$/Ar forming gas atmosphere.

The crystal structure of the grown $VO_2/RuO_2/TiO_2$ (001) heterostructures was determined by using the X-ray diffraction (XRD) (Rigaku, Ultima IV). The electronic structure of $VO_2/RuO_2/TiO_2$ heterostructure was investigated by using the soft X-ray absorption spectroscopy (sXAS) analysis, which was conducted at the Shanghai Synchrotron Radiation Facility (SSRF) on beamline BL08U1A. Electrical transport properties for the $VO_2/RuO_2/TiO_2$ (001) heterostructures were further probed using a commercial physical property measurement system (PPMS) (Quantum design), while the room-temperature material resistance were measured by using a Keithley 4200 system.

## 3. Result and discussion

Vanadium dioxide ($VO_2$) as a prototypical $3d^1$-orbital correlated oxide undergoes thermally-driven MIT behavior across the $T_{MIT}$, triggered by the splitting of $d_{//}$ orbital and/or the V-V dimerization (Figure 1a). The underlying mechanism governing the MIT in $VO_2$ remains a long-standing debate, owing to the coupled structural phase transformation in which the low-symmetry monoclinic insulating phase (space group: $P2_{1/c}$) simultaneously transforms to a rutile metallic phase (space group: $P4_{2/mnm}$). Beyond conventional semiconductors, $VO_2$ exhibits complex electronic orbital configurations that can be flexibly regulated by the following two perspectives: 1) bandwidth control by imparting the interfacial strain; 2) band-filling control using proton evolution (Figure 1b). From one aspect, imparting compressive distortion along the *c*-axis direction of rutile $VO_2$ significantly shortens the apical V-O bond length, intensifying the orbital hybridization between V-$3d$ and O-$2p$ orbitals and enlarging the bandwidth of V-3$d$ orbitals.[29] As a result, the relative stability in the metallic orbital configuration of VO$_2$ is extensively strengthened, reducing the resultant $T_{MIT}$. From the other aspect, the incorporation of hydrogens introduces the electron carriers into the conduction band of VO$_2$ that occupy the low-energy $t_{2g}$ orbital. Hydrogen-associated electron doping process triggers Mottronic phase transition from correlated electron ground state ($t_{2g}^1 e_g^0$) to electron-itinerant state ($t_{2g}^{1+A} e_g^0$) through directly altering the $d$-orbital occupation and configuration (Figure S1).[30] Therefore, interfacial strain associated with the lattice mismatch between substrate and film, together with hydrogenation, provides a powerful tuning knob for manipulating the MIT properties of correlated VO$_2$ in a controllable and reversible fashion. This advance offers considerable promise for developing tunable electronic devices across multiple disciplines by leveraging the MIT of VO$_2$.

To address the above concept, rutile RuO$_2$, with the $a$-axis lattice parameter ($a_0 = 4.49$ Å), was delicately engineered as a tunable buffer layer to precisely control the interfacial strain states in VO$_2$/TiO$_2$ (001) heterostructures. Herein, we fabricated a series of VO$_2$ (25 nm)/RuO$_2$ ($t$ nm)/TiO$_2$ (001) heterostructures with delicately controlled RuO$_2$ thicknesses ($t = 10$-$80$ nm). Noting an enlarged $a_0$ of TiO$_2$ substrate ($a_{0, sub.} = 4.59$ Å) compared with RuO$_2$ buffer layer ($a_0 = 4.49$ Å), the $in$-$plane$ tensile distortion in VO$_2$ films ($a_{0, film} = 4.54$ Å) is expected to be facilely adjusted by simply varying the thickness of RuO$_2$ buffer layer (Figure 2a). Elevating the RuO$_2$ thickness is poised to mitigate the $in$-$plane$ biaxial tensile strain in the top VO$_2$ film by suppressing the strain propagation from the lattice mismatch between the TiO$_2$ substrate and RuO$_2$ buffer layer, due to the inevitable formation of dislocations in the heterointerface region. This understanding is further confirmed by the respective X-ray diffraction (XRD) patterns (Figure 2b), where the characteristic diffraction peaks corresponding to the (002) plane of VO$_2$ and RuO$_2$ gradually shifts toward lower angles with the elevation in the RuO$_2$ buffer layer thickness (Figure S2). This result indicates an enlarged $out$-$of$-$plane$ lattice parameter for both the VO$_2$ film and RuO$_2$ buffer layer with the RuO$_2$ thickness, correlating with a progressive relaxation of $in$-$plane$ tensile strain at elevated RuO$_2$ thickness.

Further consistency in the strain regulation on the MIT functionality of VO$_2$ is demonstrated by respective temperature dependences of material resistivity ($\rho$-$T$) in Figure 2c, wherein the magnitude of $T_{MIT}$ gradually elevates with the RuO$_2$ thickness. The increase in $T_{MIT}$ is clearly evident from respective $\rho/\rho_{200\,K}$-$T$ tendencies, as the results shown in Figure S3. The $T_{MIT}$ is defined as the average value from $\rho$-$T$ tendencies during heating and cooling cycles, with individual critical temperatures determined at the local maximum of the temperature coefficient of resistance ($TCR$). It is found that the $T_{MIT}$ for the grown VO$_2$/RuO$_2$/TiO$_2$ (001) heterostructures is tunable over a wide temperature range from 297 K to 319 K (Figure 2d). Imparting an $in$-$plane$ tensile distortion enhances the orbital overlapping between V-3$d$ and O-2$p$ orbitals, thereby reducing the $T_{MIT}$ of VO$_2$. Therefore, increasing the RuO$_2$ thickness tends to suppress the $in$-$plane$ tensile strain, thereby progressively elevating the $T_{MIT}$ of VO$_2$ toward the bulk value. However, the electrical shunting through the metallic RuO₂ buffer layer inevitably degrades the transition sharpness of the grown VO₂ films, with complete metallization occurring at a thickness of 80 nm, while corresponding electrical property of the RuO₂ buffer layer are shown in Figure S4.

To probe the variations in the electronic band structure of VO₂/RuO₂/TiO₂ (001) heterostructures with varying buffer layer thicknesses, soft X-ray absorption spectroscopy (sXAS) technique is performed, as the results shown in Figure 2e. The V-$L$ edge spectrum, originating from the V $2p \rightarrow 3d$ transition, can reflect the variations in the oxidation state of vanadium. The invariant peak positions of both V-$L_{\mathrm{III}}$ and V-$L_{\mathrm{II}}$ peaks unambiguously demonstrate the retention in the vanadium valence state (i.e., +4). This finding confirms that the observed MIT modulation stems solely from interfacial strain rather than chemical doping, which cannot alter the valence state of vanadium. The compressive strain along the $c$-axis of rutile VO₂ strengthens the orbital hybridization between $d_{//}$ and $\pi^{*}$ orbitals and shortens the apical V-O bonds, stabilizing the metallic phase and reducing the $T_{\mathrm{MIT}}$. This suggests a viable strategy for adjusting the MIT property of correlated VO₂ system through bandwidth control using interfacial strain (Figure 2f).

Beyond the interfacial-strain-controlled IMT regulation, hydrogen-associated electron-doping process offers an alternative pathway for adjusting the IMT behaviors of VO₂ through band-filling control. To hydrogenate VO₂ films, a catalyst-assisted hydrogen spillover strategy was herein employed, in which dotted platinum (Pt) as catalyst significantly facilitates the dissociation of gaseous H₂ into protons and electrons (e.g., H₂ (g)→H⁺+e⁻) by markedly reducing the activation energy barrier at the gas-solid interface (Figure 3a).[31] From the perspective of structural evolution, hydrogenation induces a consistent leftward shift of characteristic diffraction peaks of VO₂ in their XRD patterns, regardless of the RuO₂ thickness (Figure 3b). This trend demonstrates an *out-of-plane* lattice expansion of VO₂ films induced by hydrogenation, primarily attributed to O-H interactions, while the *in-plane* direction is locked by the $c$-plane TiO₂ epitaxial template. In particular, hydrogen-triggered structural evolution in VO₂ is progressively depressed with an increasing thickness of the RuO₂ buffer layer (Figure 3c). This phenomenon is associated with the interfacial strain states in VO₂/RuO₂/TiO₂ (001) heterostructures, in which imparting a biaxial tensile strain extensively reduces the energy barrier for hydrogen diffusion, promoting the diffusion kinetics of hydrogens.[10] As the RuO₂ buffer layer thickness increases, the *in-plane* tensile strain originating from lattice mismatch is progressively mitigated, thereby reducing the hydrogen ion mobility and suppressing hydrogen-induced lattice expansion. The strong coupling between hydrogen-driven structural evolution and interfacial strain states bridges strain-mediated bandwidth control and hydrogenation-related band filling in correlated electron systems, bringing in a new freedom to discover novel electronic phases beyond the scope of conventional phase diagrams.

Hydrogen-triggered electronic phase transitions in VO₂ are further demonstrated by respective temperature dependences of material resistivity ($\rho$-$T$ tendency) (Figure 4a). The IMT behavior in VO$_2$/RuO$_2$/TiO$_2$ (001) heterostructure is significantly suppressed through hydrogenation, leading to a marked reduction in the transition sharpness ($\rho_{Insul.}/\rho_{Metal.}$) and facilitating a trend toward metallization. Remarkably, such the hydrogen-mediated electronic state evolution is highly reversible, in which the expected IMT functionality can be revived via exposing to the air for one month. This dehydrogenation process is related to the ultrahigh mobility of hydrogens that can be dragged out under ambient atmosphere. Nevertheless, the $T_{MIT}$ for dehydrogenated VO$_2$ is elevated compared to pristine VO$_2$, likely resulting from the partial oxidation in an oxygen-rich environment. Moreover, the transition sharpness achieved in dehydrogenated VO$_2$/RuO$_2$/TiO$_2$ (001) heterostructure via exposing to the air is lower than the pristine one, owing to the residual hydrogens remained in the deeper layer of VO$_2$ films. Such the hydrogen-triggered reversible electronic state evolution is in agreement with the structural transformation through hydrogenation, in which the characteristic diffraction peaks for dehydrogenated VO$_2$ films almost coincide with the pristine one through the exposure to the air (Figure S5).

To probe the physical origin driving hydrogen-related electronic phase modulations, the electronic band structure of VO$_2$ is characterized by using soft X-ray absorption spectroscopy (sXAS) technique in Figures 4b-4c. The V-$L$ edge spectrum, corresponding to the V $2p \rightarrow 3d$ transition, is well established as a sensitive probe of the variation in the vanadium valence state.[32] Hydrogenation renders a distinct leftward shift in both the V-$L_{III}$ and V-$L_{II}$ peaks, indicating a reduction in the oxidation state of vanadium from +4 towards +3 (Figure 4b). Such the hydrogen-triggered reduction in the vanadium valence state of VO$_2$ derived from the hydrogen-related electron doping starkly differs from the VO$_2$ under interfacial strains, in which case the changes in valence state of vanadium are not detectable. Given the presence of empty O-$2p$ states and the strong orbital hybridization of V-$3d$ and O-$2p$ orbitals, the O $1s$ spectrum provides insight into the unoccupied density of states in the conduction band of VO$_2$. Specifically, the relative intensity changes between the first and second peaks in the O $1s$ spectrum qualitatively reflect electron occupancy in the $t_{2g}$ and $e_g$ bands, respectively. The reduction in the spectral weight of the first peak relative to the second peak in the O $1s$ spectra through hydrogenation signifies an enhanced electron filling in the low-energy $t_{2g}$ band, consistent with the donation of one electron per hydrogen atom into the conduction band of VO$_2$ (Figure 4c). These synchrotron-based spectroscopic results further verify the filling-controlled Mott transition in VO$_2$ system through hydrogenation, wherein the electron carriers arising from hydrogenation preferentially occupy the low-energy $t_{2g}$ band in VO$_2$, comprising the $d_{xz}$, $d_{yz}$ and $d_{x^2-y^2}$ orbitals (Figure 4d). Consequently, hydrogenation drives the Mott transition from correlated electronic ground state in pristine VO$_2$ to an electron-itinerant state through hydrogen-related electron doping.

## 4. Conclusion

In this work, we present feasible pathways for adjusting the IMT functionality of correlated $VO_2$ through bandwidth control using interfacial strain and hydrogen-related band-filling control. Benefiting from the lattice mismatching between $TiO_2$ substrate and $RuO_2$ buffer layer, imparting an *in-plane* tensile strain to $VO_2$ film enhances the orbital overlapping between V-$3d$ and O-$2p$ orbitals and the bandwidth of $VO_2$. As a result, the $T_{MIT}$ for $VO_2$ can be flexibly manipulated across a wide temperature range of 297-319 K by simply altering the thickness of $RuO_2$ buffer layer. Alternatively, proton evolution triggers reversible Mott phase transition of $VO_2/RuO_2/TiO_2$ (001) heterostructure from correlated electron ground state based on $t_{2g}^1 e_g^0$ configuration to electron-itinerant state based on $t_{2g}^{1+\Delta} e_g^0$ configuration. In addition to driving electronic phase transition, hydrogenation renders an *out-of-plane* lattice expansion of $VO_2$ owing to O-H bonding interactions, an effect intensified by imparting an *in-plane* tensile strain. The intimate relationship between strain-mediated bandwidth control and hydrogen-related band-filling control is poised to design novel electronic phases in correlated electron systems. The present work provides new insights into rationally designing new quantum states in electron-correlated system, enabling the possibility in exploring exotic physical functionality.

**Declaration of competing interest**

The authors declare no conflict of interest.

**Acknowledgements**

This work was supported by the National Natural Science Foundation of China (No. 52401240), Fundamental Research Program of Shanxi Province (No. 202403021212123), and Scientific and Technologial Innovation Programs of Higher Education Institutions in Shanxi (No. 2024L145). The authors also acknowledge the beam line BL08U1A at the Shanghai Synchrotron Radiation Facility (SSRF) (https://cstr.cn/31124.02.SSRF.BL08U1A) and the beam line BL12B-b at the National Synchrotron Radiation Laboratory (NSRL) (https://cstr.cn/31131.02.HLS.XMCD.b) for the assistance of sXAS measurement.

**Additional information**

Supporting Information is available online or from the author.

**Author contributions**

X.Z. conceived this study, and lead the project; X.Y. and J.J. grew $VO_2$ films, and carried out the transport measurements under the supervision of X.Z.; X.Z. analyzed the results and wrote the paper; All the authors discussed the results and commented on the final manuscript.

**Correspondences:** Correspondences should be addressed: Prof. Xuanchi Zhou (*xuanchizhou@sxnu.edu.cn*).

**Figures and captions**

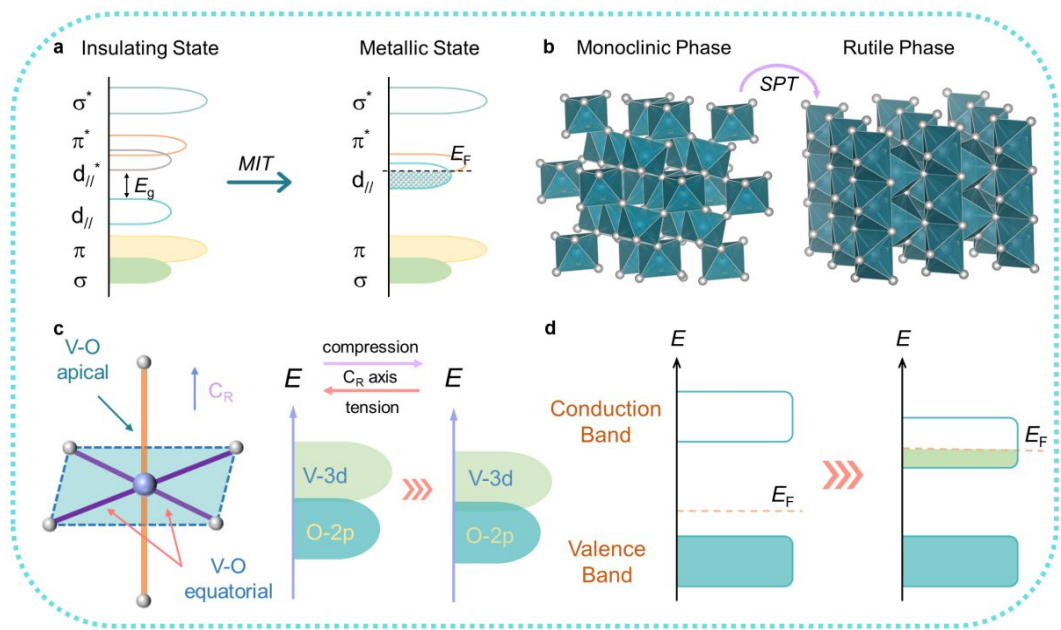

**Figure 1. a,** Schematic of the variation in the electronic band structure across the metal-insulator transition (MIT). **b,** Schematic of the structure transformation in $VO_2$ system. **c,** Schematic illustrations of the geometry configuration of $VO_6$ octahedra and corresponding electronic structure evolution in $VO_2$ system via imparting interfacial strain. **d,** Schematic of hydrogenation-mediated orbital reconfiguration via band-filling control.

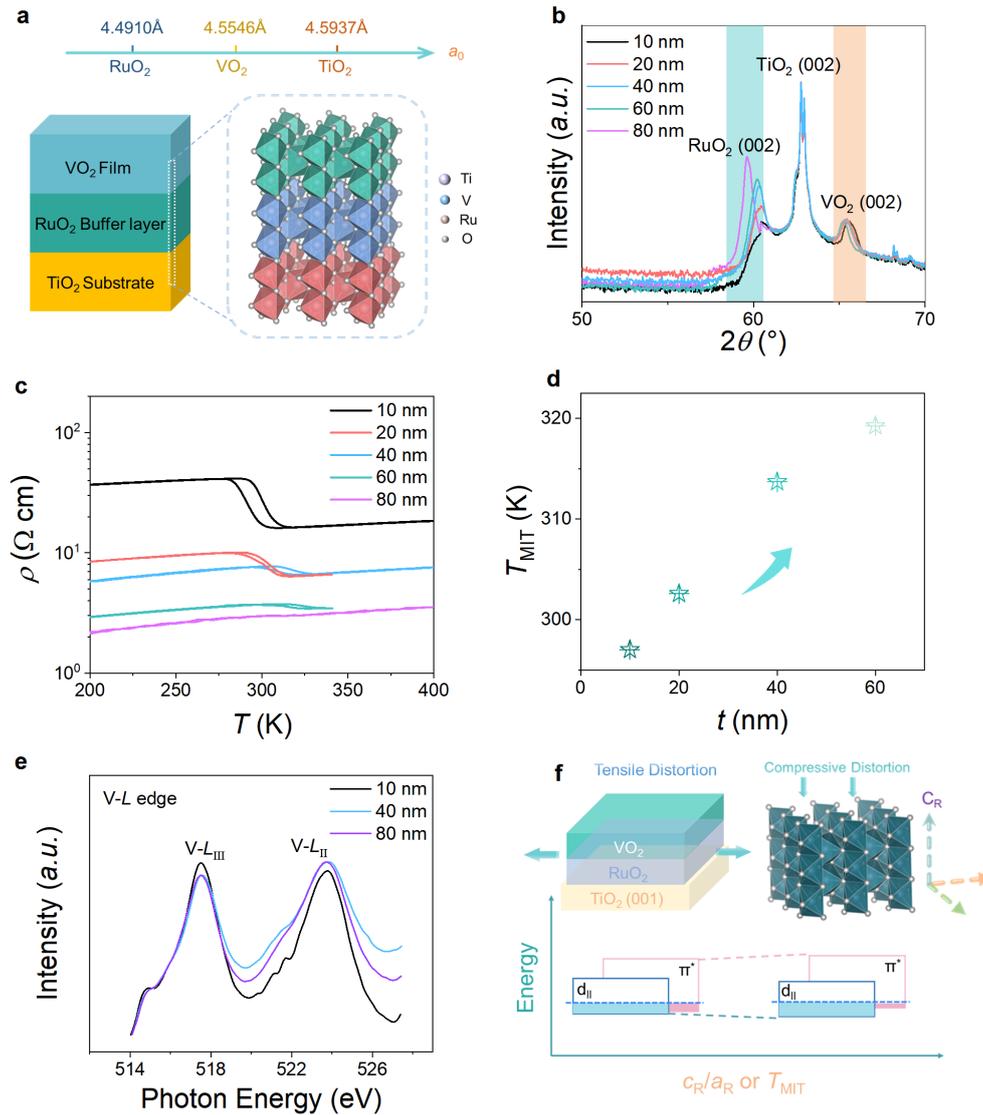

**Figure 2. a,** Schematic diagram of the $VO_2/RuO_2/TiO_2$ heterostructures. **b,** X-ray diffraction (XRD) patterns compared for $VO_2/RuO_2/TiO_2$ heterostructures with different $RuO_2$ thicknesses ranging from 10 to 80 nm. **c,** Temperature dependence of material resistivity ($\rho$-$T$) as measured for $VO_2/RuO_2/TiO_2$ heterostructures with different thicknesses. d, Schematic illustration of the relation between $RuO_2$ thickness and the transition temperature ($T_{MIT}$) in $VO_2/RuO_2/TiO_2$ heterostructures. **e,** Soft X-ray absorption spectroscopy (sXAS) compared for $VO_2/RuO_2/TiO_2$ heterostructures with different thicknesses (10 nm, 40 nm, and 80 nm). **f,** Schematic of tailoring the electronic phase transitions of $VO_2$ via manipulating the electronic orbital configuration using interfacial strains.

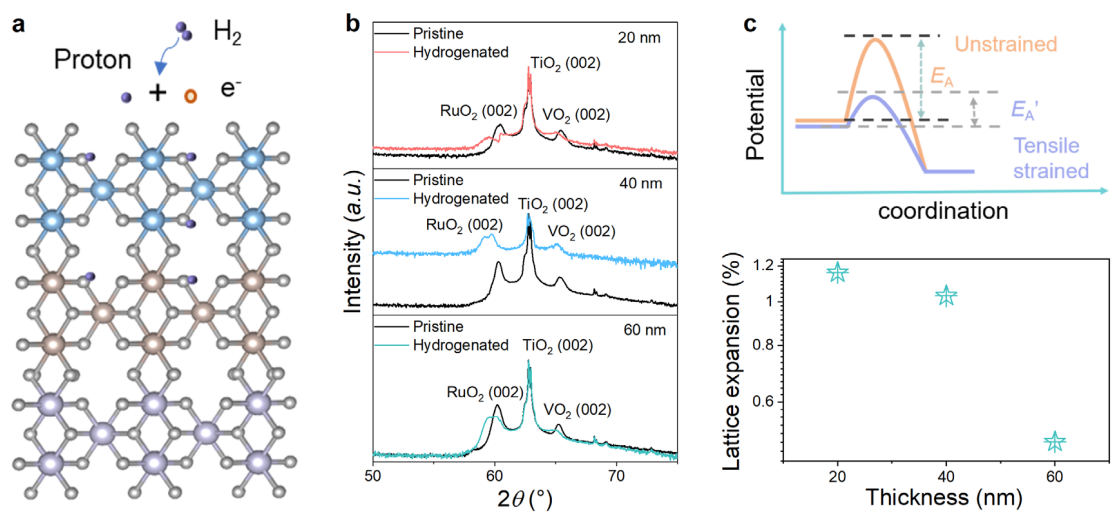

**Figure 3. a,** Schematic of hydrogenation process using Pt-assisted hydrogen spillover strategy. **b,** XRD spectra as compared for VO$_2$/RuO$_2$/TiO$_2$ heterostructures with varied RuO$_2$ thickness through hydrogenation. **c,** The hydrogen-induced lattice expansion for VO$_2$/RuO$_2$/TiO$_2$ heterostructures plotted as a function of the thicknesses of RuO$_2$ buffer layer.

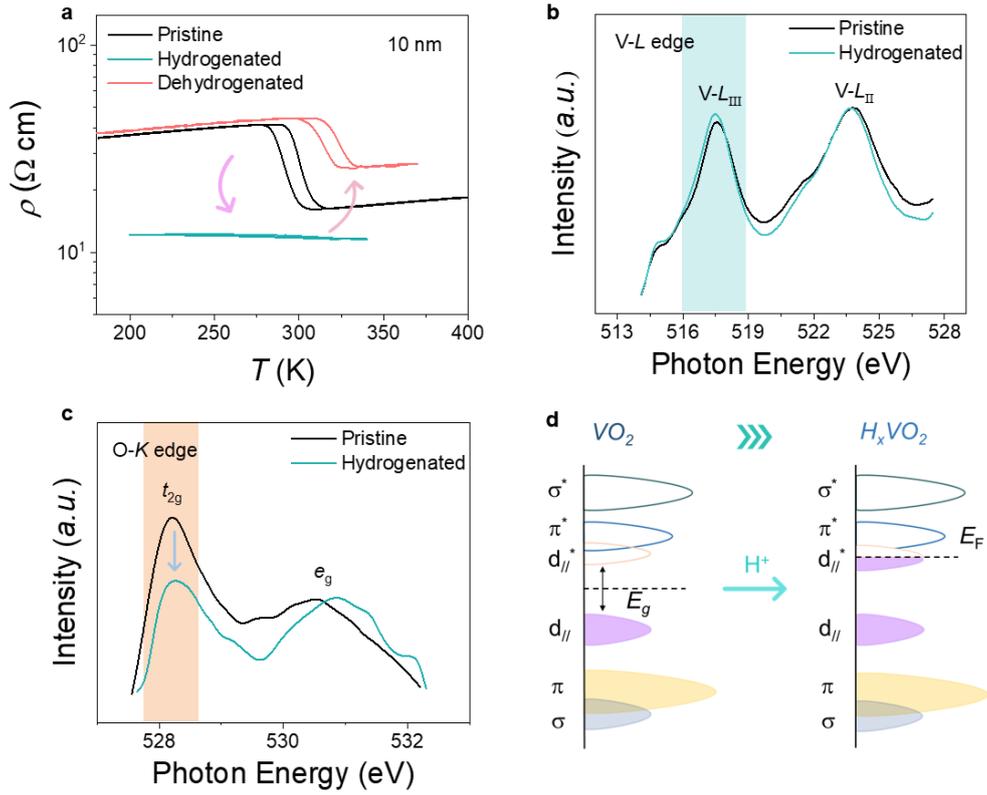

**Figure 4. a,** Temperature dependence of material resistivity ($\rho$-$T$) tendency compared for VO$_2$/RuO$_2$/TiO$_2$ heterostructures upon hydrogenation and dehydrogenation. **b-c,** sXAS spectra for **b,** V-$L$ edge and **c,** O-$K$ edge of VO$_2$/RuO$_2$/TiO$_2$ (001) heterostructures as hydrogenated at 120 °C for 3 h, in comparison with pristine state. **d,** Schematic of hydrogen-triggered electron phase modulations in correlated VO$_2$ system.